# Double-Carrier Fitting of Hall Resistance Assisted by Gate-Induced Shubnikov-de Haas Oscillations in Possible Excitonic Insulator Ta$_2$Pd$_3$Te$_5$

Xing-Chen Guo(郭兴宸)[1,2,†], An-Qi Wang(王安琪)[1,2,†], Xiu-Tong Deng(邓修同)[1,2,†], Yu-Peng Li(李宇鹏)[3], Guo-An Li(李国安)[1,2], Zhi-Yuan Zhang(张志远)[1,2], Xiao-Fan Shi(史小凡)[1,2], Xiao Deng(邓啸)[1,2], Zi-Wei Dou(窦子威)[1], Guang-Tong Liu(刘广同)[1], Fan-Ming Qu(屈凡明)[1,2], Jie Shen(沈洁)[1], Li Lu(吕力)[1,2], Zhi-Jun Wang(王志俊)[1,2], You-Guo Shi(石友国)[1,2], Hang Li(李航)[1,*], Tian Qian(钱天)[1,*]

[1]Beijing National Laboratory for Condensed Matter Physics, Institute of Physics, Chinese Academy of Sciences, Beijing 100190, China

[2]School of Physical Sciences, University of Chinese Academy of Sciences, 100049 Beijing, China

[3]Hangzhou Key Laboratory of Quantum Matter, School of Physics, Hangzhou Normal University, Hangzhou 311121, China

†These authors contributed equally to this work

*Corresponding author

e-mail: ygshi@iphy.ac.cn; hang.li@iphy.ac.cn; tqian@iphy.ac.cn

**Abstract**

Hall effect is an important phenomenon when a magnetic field is applied to materials. From the curve depicting the Hall resistance versus the magnetic field, crucial information such as carrier concentration can be extracted. If the curve exhibits a linear dependence up to rather high magnetic fields, it indicates that charge transport involves only a single type of carrier, and if a non-linear curve is measured, then the double-carrier model should be considered for fitting. However, this model involves four unknown parameters, including the concentration and mobility of the two carriers, resulting in that such fitting is usually non-unique, which significantly reduces the reliability and accuracy. In this work, a double-carrier platform was constructed on a probable excitonic insulator $Ta_2Pd_3Te_5$, and the four-parameter fitting based on the double-carrier model was simplified to a single-parameter fitting by employing methods such as analyzing the shape of the Hall resistance curve and generating gate-induced Shubnikov-de Haas oscillations. Thus, we provide a reliable method for double-carrier fitting of Hall resistance and a new evidence for the existence of excitonic-insulator state in $Ta_2Pd_3Te_5$.

**Introduction**

Hall effect[1] is an complex and extensive system that plays an important role in condensed matter physics. Together with anomalous Hall effect[2], quantum Hall effect[3,4], fractional quantum Hall effect[5], quantum anomalous Hall effect[6,7], quantum spin Hall effect[8,9] and so on, Hall effect exhibits a great number of novel phenomena, and witnesses the development of Topological physics[10-13] and even Majorana physics[14,15]. Among the numerous practical applications of Hall effect, calculating the carrier concentration of materials is a widely used purpose.

When there is only a single type of carriers in material, the Hall resistance will linearly increase as the magnetic field strengthen, as described by

$$R_{Hall} = \frac{1}{ne} B \quad (1)$$

where $R_{Hall}$ is the Hall resistance, $B$ is the magnetic induction intensity, $e$ is the charge constant, and $n$ is 2-dimensional (2D) concentration of charge carrier, here we specify that if the carrier is electron, $n$ takes negative values, otherwise if the carrier is hole, *n* takes positive values. In this case, positive charge carriers cause positive slope and negative charge carriers cause negative slope. Fitting the $R_{Hall}$-*B* curve and extracting its slope, then the 2D carrier concentration *n* can be calculated.

When there are two types of carriers are involved in charge transport, the $R_{Hall}$-*B* curve will be non-linear. The double-carrier model[16,17] is proposed to illustrate such situation,

$$R_{Hall}=\frac{B}{e}\frac{n_1\mu_1^2+n_2\mu_2^2+(n_1+n_2)\mu_1^2\mu_2^2 B^2}{(n_1\mu_1+n_2\mu_2)^2+(n_1+n_2)^2\mu_1^2\mu_2^2 B^2} \quad (2)$$

where $n_1$, $\mu_1$, $n_2$, $\mu_2$ are the 2D concentration and mobility of the two types of charge carriers respectively, and given that the sign of $n_1$ and $n_2$ represents the sign of carriers, $\mu_1$ and $\mu_2$ should also keep the same sign with carriers to guarantee that the equation is valid. However, if equation (2) is used directly to fit the double-carrier $R_{Hall}$-*B* curve, many groups of $n_1$, $\mu_1$, $n_2$ and $\mu_2$ can be suitable, for such a four-parameter fitting is usually non-unique, which significantly reduces the reliability and accuracy of double-carrier model fitting. In some situations the magnetoresistance (MR) curve can be used to constrain the range of the fitting parameters[16,17],

$$MR=\frac{n_1 n_2 \mu_1 \mu_2 (\mu_1-\mu_2)^2 B^2}{(n_1\mu_1+n_2\mu_2)^2+(n_1\mu_1^2+n_2\mu_2^2)^2 B^2} \quad (3)$$

but in a great number of cases, the MR behavior does not conform to equation (3). For instance, in topological insulators[18] (TI) or materials with impurities[19], MR can exhibit a linear dependence on *B*, moreover, weak localization[20-24], weak anti-localization[25] and giant magneto-resistance effect[26,27] will also drive MR deviate from equation (3). Thus, developing an accurate and reliable method for double-carrier model fitting becomes an urgent question to be solved.

In this work, we realized a new double-carrier model fitting method on a probable excitonic insulator[28,29] (EI) $Ta_2Pd_3Te_5$, which has been revealed as a sophisticated quasi-one-dimensional (Q1D) transition-metal telluride hosting a variety of novel physical phenomena, including anisotropic tunable Luttinger liquid in edge state[30,31], second-order topological insulators[32], quantum-spin-Hall states[29] and edge supercurrents[33], furthermore, it can also act as a topological thermometer spanning from millikelvin temperatures to room temperatures[34]. We fabricated a $Ta_2Pd_3Te_5$ few-layer device with a graphite/hBN back gate. On the one hand, for the fact that EI is an insulator whose band gap is opened by Coulomb interaction between electron and hole[35-37], it is supposed to be easy to find two types of charge carriers, more precisely, electron and hole in an EI. On the other hand, a graphite back gate electrode can induce Shubnikov-de Haas

(SdH) oscillation[38], which is proved to be an important support for the double-carrier model fitting.

**Method**

To simplify the process of fitting, one strategy is to reduce the number of the unknown parameters, and experimentally, substituting measurable variables for certain components in the equation (2) should be an effective approach.

It is not hard to find that in low-field limit, $R_{Hall}$-$B$ curve is linear,

$$R_{Hall}=\frac{B}{e}\frac{n_1\mu_1^2+n_2\mu_2^2}{(n_1\mu_1+n_2\mu_2)^2} \quad (4)$$

and its slope

$$K=e\rho_{2D}^2(n_1\mu_1^2+n_2\mu_2^2) \quad (5)$$

where $\rho_{2D}$ is the 2D resistivity at zero field, which is defined as

$$\rho_{2D}=R_0\frac{W}{L} \quad (6)$$

and $R_0$ is the resistance at zero field, $W$ and $L$ are the width and length of the device. Thus,

$$\rho_{2D}=\frac{1}{e(n_1\mu_1+n_2\mu_2)} \quad (7)$$

And evidently, $R_{Hall}$-$B$ curve is also linear in high-field limit,

$$R_{Hall}=\frac{1}{(n_1+n_2)e}B \quad (8)$$

However, in many cases the high-field limit is too large for laboratories to reach, so here we only speculate the sign of its slope, which represents the sign of the total carrier concentration, that is, which type of carrier dominates, from the shape of the curve. And specially, if the system is at charge neutral point (CNP), that is, there are holes and electrons involved in charge transport and their concentrations are the same, namely $n_1=-n_2=n$, the $R_{Hall}$-$B$ curve will be linear from low-field limit to high-field limit, just like a single-carrier model, but its slope is

$$K_{n_1=-n_2}=\frac{1}{ne}\frac{\mu_1+\mu_2}{\mu_1-\mu_2} \quad (9)$$

Setting $N=n_1+n_2$, $U=\mu_1\mu_2$ and substituting equation (5) and (7) into equation (2), we get

$$R_{Hall}=B\frac{K+e\rho_{2D}^2NU^2B^2}{1+e^2N^2U^2B^2} \quad (10)$$

where $K$ and $\rho_{2D}$ are both measurable variables, if $N$ can be measured as well, $U$ will be the only unknown fitting parameter in the double-carrier model, and can be fitted precisely. Thus $n_1$, $\mu_1$, $n_2$, $\mu_2$ can be solved from $N=n_1+n_2$, $U=\mu_1\mu_2$, equation (5) and equation (7).

Actually the total carrier concentration $N$ is indeed measurable. When high-field limit is reachable, equation (8) provides an convenient way to fit $N$. Whereas when high-field limit is unreachable, gating is another way to calculate it. For 2D metal devices, the variation in 2D carrier concentration $\Delta N$ caused by gating can be deduced by

$$e\Delta N = \Delta V_G C_G \qquad (11)$$

where $\Delta V_G$ is the change of gate voltage, and $C_G$ is the electric capacity on per unit area and

$$C_G = \frac{\varepsilon_0 \varepsilon_r}{d} \qquad (12)$$

where $\varepsilon_0$ and $\varepsilon_r$ are vacuum/relative dielectric constant, and $d$ is the thickness of the dielectric layer. The CNP can be located by whether the $R_{Hall}$-$B$ curve is linear throughout the entire range of magnetic field strengths.

However, for an insulator or semiconductor with band gap, the effect of gating is manifested not merely in the change of carrier concentration, but also in the variation of the chemical potential[39] $\Delta u$,

$$e\Delta V_G = \Delta u + \frac{e^2 \Delta N}{C_G} \qquad (13)$$

It shows that calculating the effective gating voltage should isolate the influence of chemical potential variation. So deducing carrier concentration through gate voltage directly will introduce measurement errors.

It is universally acknowledged that graphite can exhibit SdH oscillation in a varying magnetic field[38,40], thus, when graphite acts as the gate electrode in a device, the device capacitance will be modulated in synchrony, and consequently, the carrier concentration as well. As a result, the resistance of the device will also show a SdH oscillation due to electrostatic sample-gate coupling[38], and the frequency $f$ of the gate-induced SdH oscillation reflects the Fermi surface area of the graphite gate[41],

$$S_F = \pi k_F^2 = \frac{2\pi e}{\hbar} f \qquad (14)$$

where $k_F$ is the Fermi wave vector. Hence, considering the two-dimensional characteristic of graphite, the 2D carrier concentration $n_{2D}$ of the graphite gate can be known by

$$n_{2D} = \frac{k_F^2}{2\pi} = \frac{2e}{h} f \qquad (15)$$

Furthermore, the sample layer should keep the same variety in 2D carrier concentration with gate electrode. As the CNP is located by the shape of $R_{Hall}$-$B$ curve, the total 2D carrier concentration of the sample layer can be calculated easily.

**Results**

Ta$_2$Pd$_3$Te$_5$ is a layered van der Waals (vdW) material and each of its layers is constructed of Ta$_2$Te$_5$ chains and Pd atoms (Fig. 1(a)). We fabricated Ta$_2$Pd$_3$Te$_5$ devices with graphite or Ti/Au back gate, labeled as #1 and #2 respectively, as shown in the inset of Fig. 2(a) and (c). We measured the temperature ($T$) dependent resistance ($R$) curve of device #1 and extracted its varying band gap against temperature (Fig. 1(b)). In high temperature regime, gap decreases while warming, which consist with the characteristic of EI[28]. Then we confirmed that there are two types of carriers in this device by the shape of $R_{Hall}$-$B$ curve shown in Fig. 1(d). and Fig. 1(c) shows that back gate voltage ($V_{bg}$) can tune the Fermi-level from its valence band to conduction band crossing the CNP.

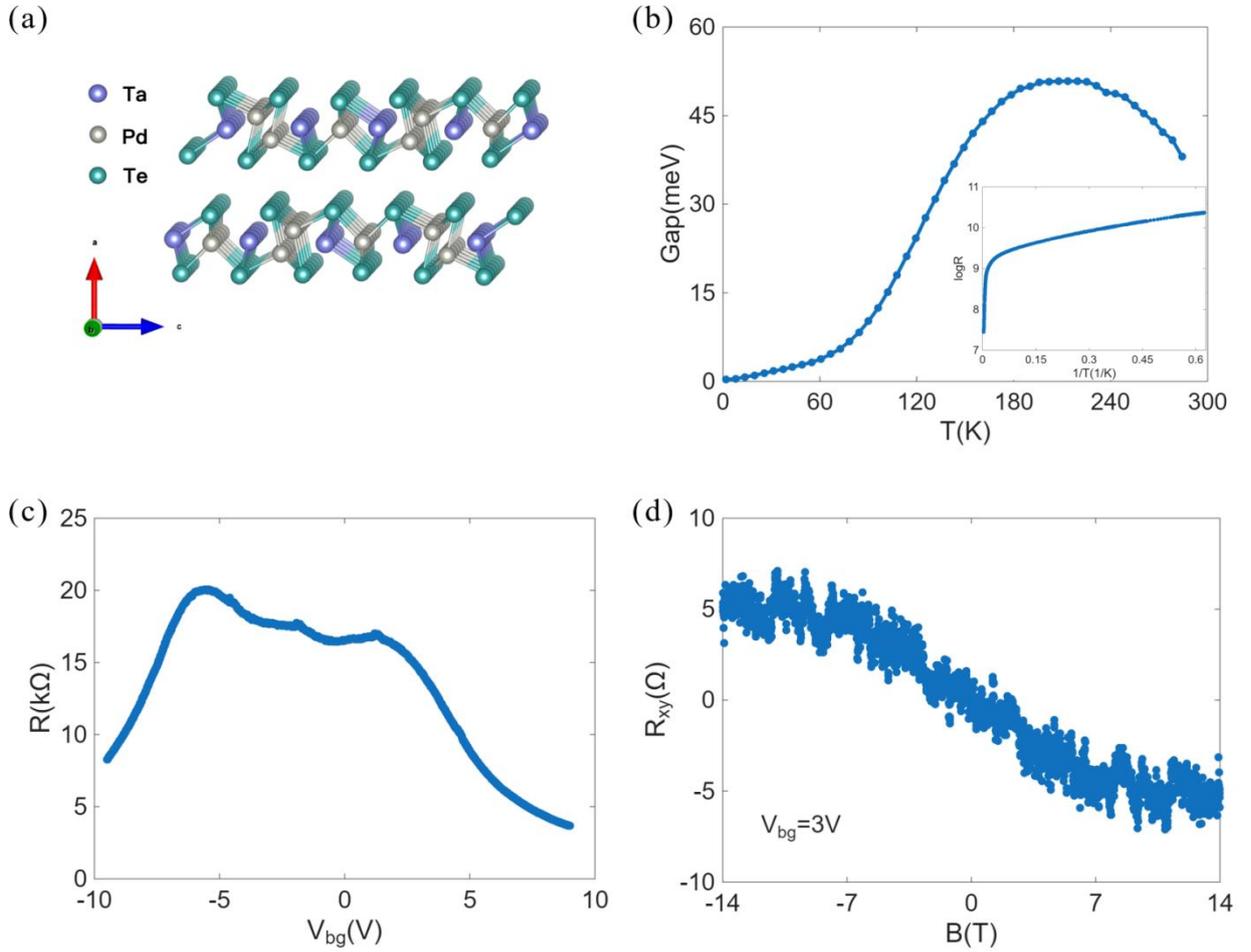

Fig. 1. Fundamental properties of Ta$_2$Pd$_3$Te$_5$. (a) Crystal structure of Ta$_2$Pd$_3$Te$_5$. (b) Plot of band gap versus temperature. Inset: resistance versus temperature plotted by log$R$ versus 1/$T$. Band gap in (b) is extracted from the log$R$-1/$T$ curve by $\Delta = 2k_B \cdot d(\log R)/d(1/T)$, where $k_B$ is Boltzmann constant. (c) Plot of resistance versus back gate voltage at $T$=5K. (d) Plot of Hall resistance versus magnetic field B at $T$=5K and $V_{bg}$=3V.

Then we examined the gate-induced SdH oscillation in $Ta_2Pd_3Te_5$. Applying magnetic field on both device #1 and #2, as Fig. 2(a) and (c) shown, it is evident to see that device #1 with graphite gate exhibits quantum oscillation while device #2 with Ti/Au gate exhibits no distinct oscillation at all. Subtracting the background, the oscillation periods of device #1 are found to be constant when plotted against $1/B$, and the frequency can be extracted by fast Fourier transform (FFT) as 72.8T, while device #2 exhibits only noise of measurement, as shown in Fig. 2(b) and (d). This indicates that the quantum oscillation is totally induced by the graphite gate, but has nothing to do with intrinsic properties of $Ta_2Pd_3Te_5$.

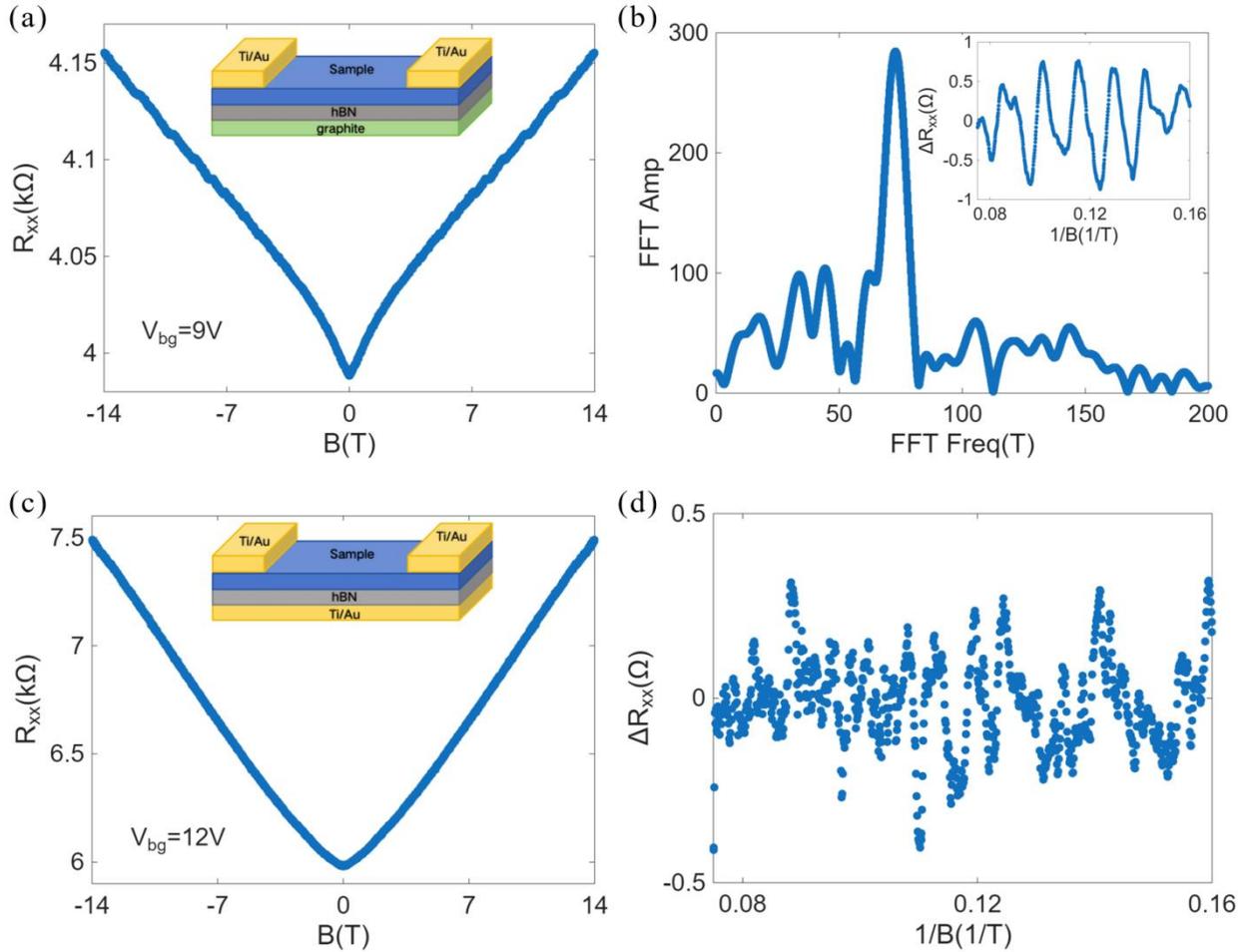

Fig. 2. Quantum oscillation induced by graphite gate. (a) Plot of the longitude resistance $R_{xx}$ versus magnetic field at $T$=2K and $V_{bg}$=9V. Inset: schematic of device #1. (b) FFT plot of the quantum oscillation in (a). Inset: the quantum oscillation extracted from (a) plotted as the oscillation amplitude versus $1/B$. (c) Plot of the longitude resistance $R_{xx}$ versus magnetic field at $T$=2K and $V_{bg}$=12V. Inset: schematic of device #2. (d) the quantum oscillation extracted from (c) plotted as the oscillation amplitude versus $1/B$.

Furthermore, we measured the SdH oscillation at a series of $V_{bg}$ from -9.6V to 9.0V and plot the data as waterfall diagram and heat map, as shown in Fig. 3(a) and (b). The peaks or valleys of the oscillations can be connected to form straight lines that converge at the origin and construct Landau fans, which perfectly matches the band structure of graphite. Thus the 2D carrier concentration in the graphite gate layer can be calculated by equation (15), as shown in Fig. 3(c), $n_{2D}$ does not vary strictly linearly with $V_{bg}$, which confirms that for semiconductors or insulators with band gap, deducing carrier concentration through gate voltage directly will introduce measurement errors. And according to the characteristic of SdH oscillation, the location of peaks do not vary with temperature, as shown in Fig. 3(d), indicating that the tunability of gate do not change despite of warming or cooling.

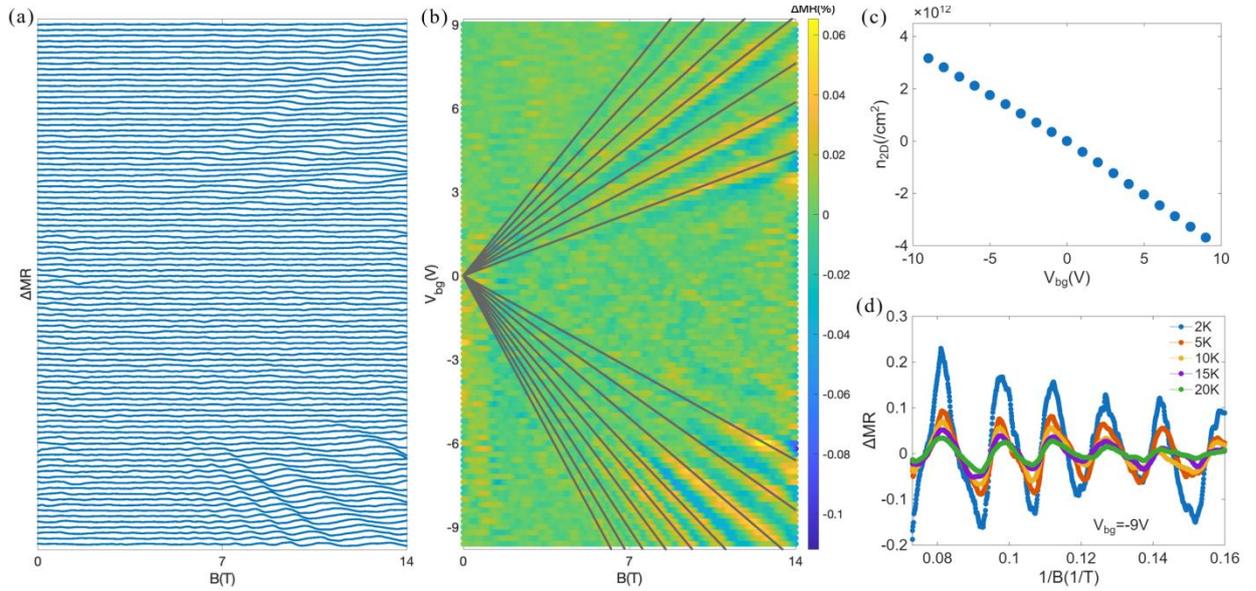

Fig. 3. Other properties of the gate-induced SdH oscillation. (a)(b) Waterfall plot and heat map of the oscillation amplitude in MR versus magnetic field at different $V_{bg}$ and $T$=20K. Measurements are taken from $V_{bg}$=-9.6V to 9.0V at intervals of 0.2V. Gray lines in (b) connect the peaks of oscillations. (c) Plot of the 2D carrier concentration in the graphite gate layer versus $V_{bg}$ calculated through SdH oscillations. (d) Plot of the oscillation amplitude in MR versus $1/B$ at different temperature and $V_{bg}$=-9V.

The next step is to find the CNP. As the $R_{Hall}$-$B$ curve at different $V_{bg}$ illustrated (Supplementary Fig. 1), the slope at high magnetic field when $V_{bg}$<4V can be implied to be positive while that when $V_{bg}$>4V can be implied to be negative, and the $R_{Hall}$-$B$ curve at $V_{bg}$=4V is fitted to be a straight line within the range of

$|B|<14T$. It indicates that the CNP is at $V_{bg}=4V$, hence the total 2D carrier concentration $N$ in the sample layer can be calculated through the 2D carrier concentration $n_{2D}$ in the graphite gate layer by $N=n_{2D}-n_{2D}(V_{bg}=4V)$. After fitting and solving through the method above, we get the 2D carrier concentration and mobility of hole and electron at different $V_{bg}$ in this device, as shown in Fig. 4, and the fitting results are illustrated as the red lines in Supplementary Fig. 1.

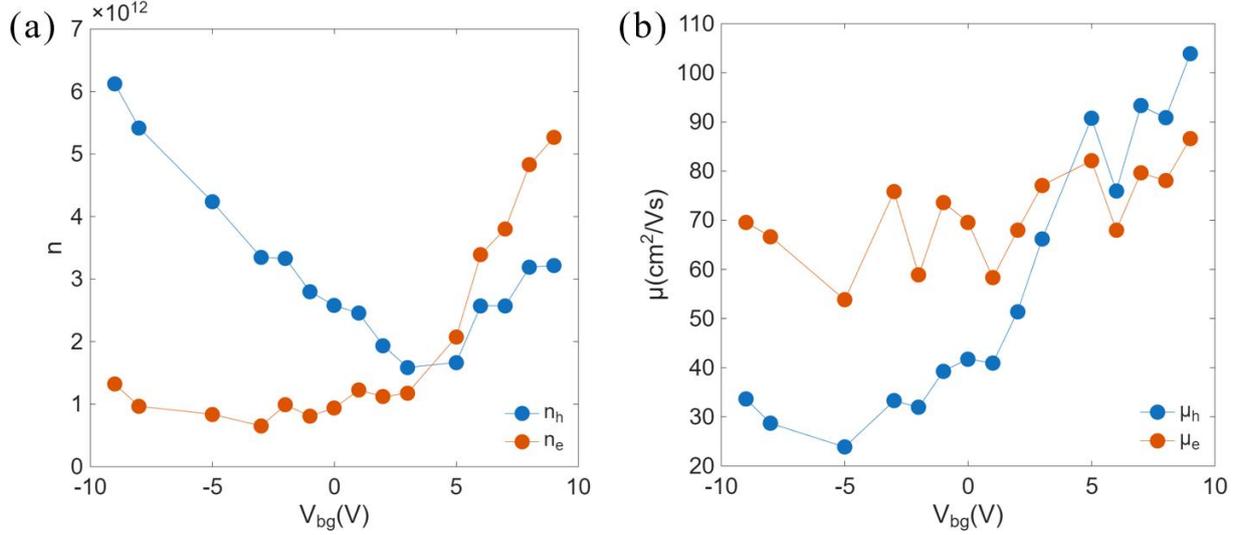

Fig. 4. Fitting results of $R_{Hall}$-$B$ curves at different $V_{bg}$ on device #1. (a) Plot of 2D carrier concentration of hole and electron versus $V_{bg}$. (b) Plot of mobility of hole and electron versus $V_{bg}$.

**Discussion**

A closer examination of Fig. 4 reveals two anomalies: first, the resistance at the CNP is not the maximum; second, there is an anomalous increase in hole concentration on the right side of the CNP, for comparison, on the left side of CNP, the electron concentration remains nearly constant. We attribute both anomalies to the probable breakdown of excitonic insulator[42]. Fig. 4(b) shows that the hole mobility exhibits a sharp increase with rising gate voltage, which indicates a breakdown of excitonic insulator - the electron-hole pairs are broken apart and the bound charge carriers become activated and released, leading to an anomalous increase in hole concentration on the right side of the CNP and a consequent reduction in resistance.

**Summary**

In this work, we developed a new method for double-carrier model Hall resistance fitting and provide a new approach for carrier concentration measurement by gate-induced SdH oscillation. By combining the two methods, we are able not only to accurately measure the carrier concentration in two-dimensional semiconductor or insulator devices without being affected by errors induced by changes in the Fermi surface, but also, when fitting the Hall resistance for double-carrier systems, replace the unknown components in the fitting formula with measurable quantities, which reduces the number of fitting parameters from four to one, thereby significantly enhancing the accuracy and reliability of the fitting process. We realized these two methods on a $Ta_2Pd_3Te_5$ device and calculated the carrier concentration and mobility of the two type of carriers at different back gate voltages, and discovered anomalies in carrier concentration, mobility and resistance, which can be important evidences for the existence of excitonic insulator breakdown.

**Acknowledgement**

The work of J.S., L.L., F.Q. and G.L. was supported by the National Key Research and Development Program of China (Grant No. 2023YFA1607400, Grant No. 2024YFA1613200), the Beijing Natural Science Foundation (Grant No. JQ23022), the National Natural Science Foundation of China (Grant Nos. 12174430 and 92365302), and the Synergetic Extreme Condition User Facility (SECUF; https://cstr.cn/31123.02.SECUF). Y.S. acknowledges support from the National Key Research and Development Program of China ( No. 2024YFA140840), the National Natural Science Foundation of China (No. U22A6005). Y.L. acknowledges support from the National Natural Science Foundation of China (Grant No. 12404154). The work of Z.D. was supported by the National Natural Science Foundation of China (Grant No. 12504561). The work of the other authors was supported by the National Key Research and Development Program of China (Grant Nos. 2019YFA0308000, 2022YFA1403800, 2023YFA1406500, and 2024YFA1408400), the National Natural Science Foundation of China (Grant Nos. 12274436, and 12274459), the Beijing Natural Science Foundation (Grant No. Z200005). The work is also funded by the Chinese Academy of Sciences President's International Fellowship Initiative (Grant No. 2024PG0003).

# Supplementary for

# Double-Carrier Fitting of Hall Resistance Assisted by Gate-Induced Shubnikov-de Haas Oscillations in Possible Excitonic Insulator Ta$_2$Pd$_3$Te$_5$

Xing-Chen Guo(郭兴宸)[1,2,†], An-Qi Wang(王安琪)[1,2,†], Xiu-Tong Deng(邓修同)[1,2,†], Yu-Peng Li(李宇鹏)[3], Guo-An Li(李国安)[1,2], Zhi-Yuan Zhang(张志远)[1,2], Xiao-Fan Shi(史小凡)[1,2], Xiao Deng(邓啸)[1,2], Zi-Wei Dou(窦子威)[1], Guang-Tong Liu(刘广同)[1], Fan-Ming Qu(屈凡明)[1,2], Jie Shen(沈洁)[1], Li Lu(吕力)[1,2], Zhi-Jun Wang(王志俊)[1,2], You-Guo Shi(石友国)[1,2], Hang Li(李航)[1,*], Tian Qian(钱天)[1,*]

[1]Beijing National Laboratory for Condensed Matter Physics, Institute of Physics, Chinese Academy of Sciences, Beijing 100190, China

[2]School of Physical Sciences, University of Chinese Academy of Sciences, 100049 Beijing, China

[3]Hangzhou Key Laboratory of Quantum Matter, School of Physics, Hangzhou Normal University, Hangzhou 311121, China

†These authors contributed equally to this work

**\*Corresponding author**

e-mail: ygshi@iphy.ac.cn; hang.li@iphy.ac.cn; tqian@iphy.ac.cn

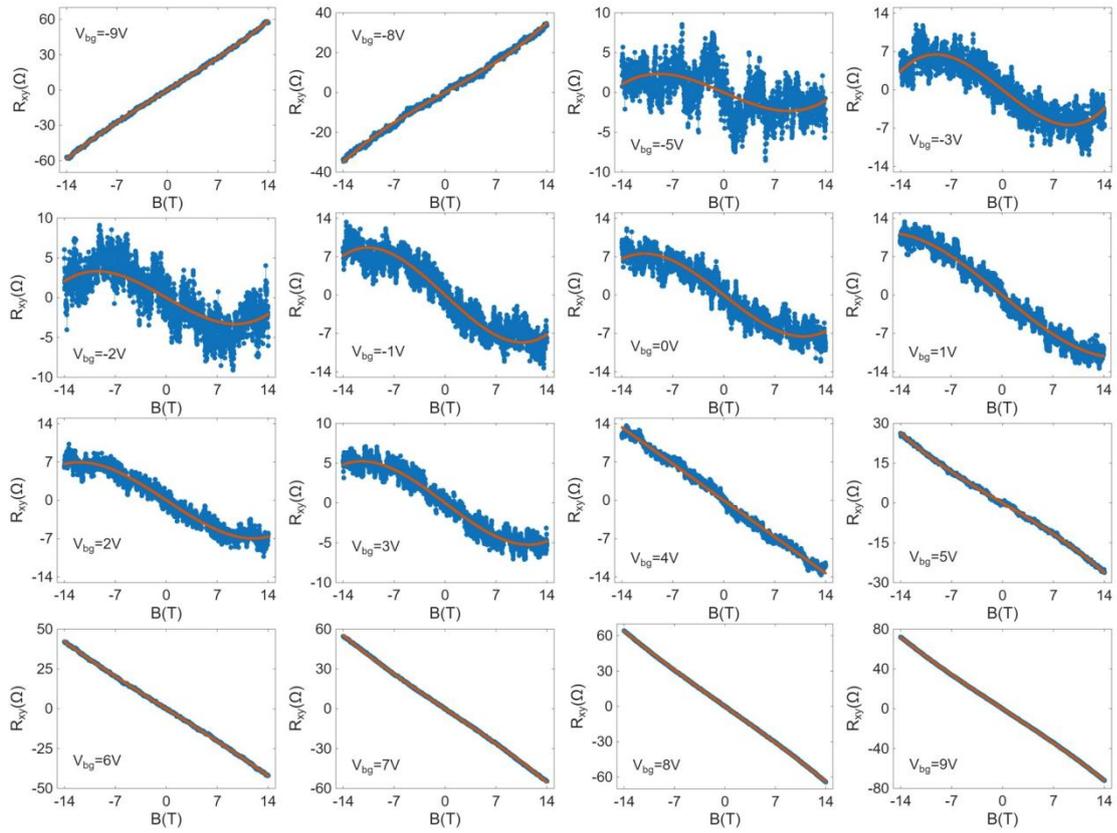

**Supplementary Fig. 1** Plot of Hall resistance versus magnetic field at different back gate voltages and $T$=5K on device #1. Blue dots are experimental data and red lines are fitting results.